\documentstyle[epsfig,amssymb]{elsart}

\newcommand{\dedx}{${\d}E/{\d}x$}

\begin{document}

\begin{frontmatter}

\title{A Compact Solid State Detector for Small Angle Particle Tracking}

\author[unipv,infnpv]{S. Altieri},
\author[unipv]{O. Barnaba},
\author[infnpv]{A. Braghieri\thanksref{email}},
\author[unipv,infnpv]{M. Cambiaghi},
\author[infnpv]{A. Lanza},
\author[infnpv]{T. Locatelli},
\author[unipv,infnpv]{A. Panzeri},
\author[infnpv]{P. Pedroni},
\author[unipv,infnpv]{T. Pinelli}
\address[unipv]{Dipartimento di Fisica Nucleare e Teorica,
 Universit\`a degli Studi di Pavia, via Bassi 6, 27100 Pavia, Italy}
\address[infnpv]{INFN-Sezione di Pavia, via Bassi 6, 27100 Pavia, Italy}

\author{P. Jennewein},
\author{M. Lang},
\author{I. Preobrazhensky}
\address{Institut f\"ur Kernphysik, Universit\"at Mainz,
J.J. Becher Weg 45, D 55099 Mainz, Germany}

\author{J.R.M. Annand},
\author{F. Sadiq}
\address{Department of Physics and Astronomy, University of Glasgow, 
Glasgow G12 8QQ, Scotland}

\thanks[email]{Corresponding author. E-mail address: 
braghieri@pv.infn.it}

\begin{abstract}

MIDAS (MIcrostrip Detector Array System) is a compact silicon 
tracking telescope for charged particles emitted at small angles in
intermediate energy photonuclear reactions.

It was realized to increase the angular acceptance of
the DAPHNE detector and used 
in an experimental program
to check the Gerasimov-Drell-Hearn sum rule at the 
Mainz electron microtron, MAMI.

MIDAS provides a trigger for charged hadrons,
p/$\pi^\pm$ identification and particle tracking in the
region $7^\circ < \vartheta < 16^\circ$.

In this paper we present the main characteristics of MIDAS 
and its measured performances.

\end{abstract}

\end{frontmatter}

\section{Introduction}

An experimental program has been started at the MAMI tagged photon beam
facility to verify the 
Gerasimov-Drell-Hearn (GDH) sum rule. A detailed description of the
physical motivations are reported in 
refs.~\cite{gdhprop1,gdhprop2,gdhprop3,gdhprop4}.
The experiment requires the measurement of
the total absorption cross section of circularly polarized photons
on longitudinally polarized nucleons
over a wide range 
of energy. The GDH sum rule is related to the difference
of the two helicity dependent cross sections, considering
target and beam spins parallel and antiparallel.

As a consequence a large angular and momentum acceptance detector 
is needed
to reduce extrapolations and then minimize systematic errors in
the total cross section evaluation.

The DAPHNE detector \cite{daphne} already meets most of the above
requirement, covering 94\% of the total solid angle
($21^\circ \le \vartheta \le 159^\circ$;
$0^\circ \le \varphi \le 360^\circ$). 
However in order to increase the acceptance in the angular region 
$\theta < 21^\circ$
an upgrade of the detector was necessary.
Since the DAPHNE structure includes a mechanical frame 
which shadows particle transmission in the region
$5^\circ < \vartheta < 21^\circ$,
the extension of the acceptance into this range
required a set-up inside the frame.
For angles $\vartheta < 5^\circ$ beam-halo interference
is a problem unless the detector is placed far downstream of
the target.

\begin{figure}[hbt]
  \begin{center}
    \epsfig{file=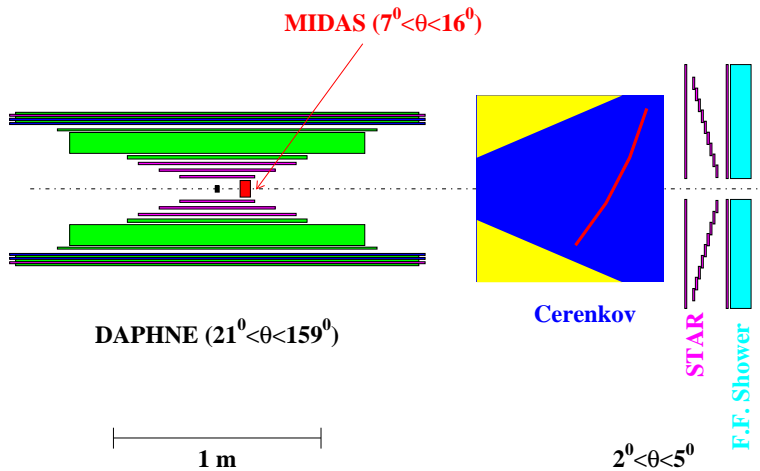, width=10cm}
  \end{center}
  \caption{Layout of the GDH set-up}
  \label{gdh}
\end{figure}

Thus the forward-angle system has two components:
MIDAS, described here and in refs. \cite{pale,midas1},
covers the region $7^\circ < \vartheta < 16^\circ$
and is placed
deep inside the DAPHNE frame, close to the target, while
the STAR detector and a scintillator-lead device \cite{star} are placed 
externally to the frame covering the region 
$2^\circ < \vartheta < 5^\circ$.
Between the two devices an aerogel and N$_2$ gas \v{C}erenkov
detector enables
electron detection with a threshold of about 15 MeV and
an efficiency $>$99\%. 
Its anti-coincident signal is used to
suppress electromagnetic background from the trigger.
The full experimental set-up is shown in Figure~\ref{gdh}.

\section{Detector design}

The volume around the target was very limited, so
we decided to use silicon detectors in order to obtain
a compact geometry for MIDAS.

\begin{figure}[hbt]
  \begin{center}
    \epsfig{file=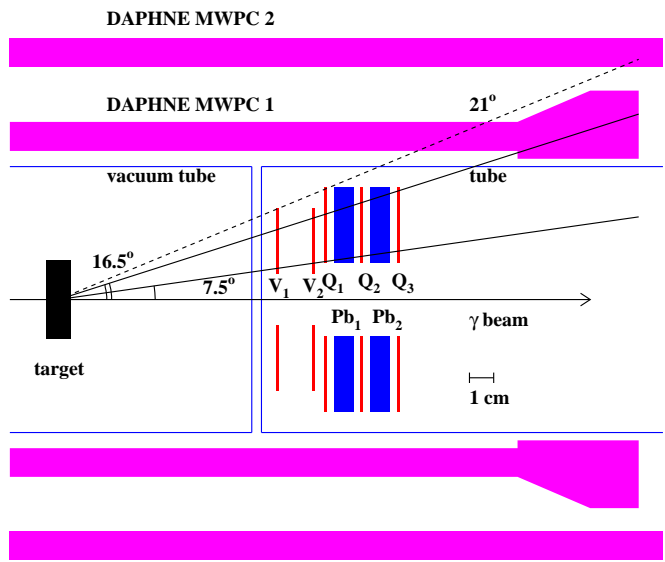, width=9cm}
  \end{center}
  \caption{Schematic side view of MIDAS}
  \label{midas}
\end{figure}

MIDAS includes two parts (see Figure~\ref{midas}):
a tracking section consisting of two annular double sided silicon 
detectors
(V$_1$ and V$_2$) and an annular silicon/lead sandwich 
(Q$_1$, Pb, Q$_2$, Pb, Q$_3$) for energy measurement.

A central hole allows  the primary $\gamma$ beam to go through.

The p-side of the tracking detectors is divided 
into 48 concentric rings, while 
the n-side is segmented into 16 radial sectors.
The signals are carried out by a
flexi-rigid cable consisting of 5 overlapping layers each screened
by a grounded wire-netting structure.

The remaining
detectors Q$_1$, Q$_2$ and Q$_3$ are single sided with the p-side
segmented into 4 radial sectors (quadrants).

All detectors were manufactured by Micron Semiconductor Ltd, UK.
Their main geometrical parameters are listed in table~\ref{tab1}.

\begin{table}[hbt]
  \caption{Geometrical characteristics of the silicon detectors}
  \medskip
  \begin{center}
    \begin{tabular}{ccc}
      \hline
      & \bf{Tracking Detectors} & \bf{Quadrant Detectors}\\\hline\hline
      \bf{Thickness} & 1000 $\mu$m & 1000 $\mu$m \\ \hline
      \bf{Geometry} & Annular Ring & Annular Ring \\
      inner radius & 11 mm & 14 mm \\
      outer radius & 35 mm & 43 mm \\ \hline
      \bf{Type} & Double sided & Single sided\\
      p side & 48 rings & 4 sectors \\
      n side & 16 sectors & - \\ \hline
     \end{tabular}
  \end{center}
  \label{tab1}
\end{table}

The detectors and the lead absorbers were mounted
inside an Aluminum tube fitted into the forward hole of the DAPHNE 
detector.
The detector signal cables exit from the downstream opening of the tube
leaving the $\vartheta < 5^\circ$ region free
for the particle transmission towards the other forward devices.

The tube was closed at both ends by 25~$\mu$m thick Aluminum windows
and supplied with an Argon flow of a few $\ell$/h to ensure that
the detectors current is stable and does not depend
on atmospheric impurities or humidity.

\section{Principle of operations}

GEANT simulations were used to optimize the geometrical set-up
by taking into account the experimental conditions at MAMI energy.
MIDAS provides the following functions:

\begin{enumerate}
\item charged-hadron triggering;
\item charged-particle tracking;
\item discrimination between e$^\pm$, $\pi^\pm$ and p
\end{enumerate}

\subsection{Hadron trigger}\label{ch:ht}

Two trigger conditions were implemented to detect protons and 
pions with the minimum background contamination.

The most important source of
background is due to electrons (and positrons) from pair 
production and
Compton scattering processes occurring in the experimental target.
The double lead sandwich, whose
total thickness corresponds to $\simeq$3 radiation lengths,
was designed to absorb most of these electrons.

On the contrary, almost all the pions and high energy protons
at forward angles
are completely transmitted, giving a signal from all detectors.
Therefore a good trigger for high energy 
hadrons was obtained by the coincidence of the 3 
quadrant detectors Q$_1$, Q$_2$, Q$_3$.

The low energy protons stopped inside MIDAS produce
a Q$_1$-Q$_2$ coincidence. These particles have a large energy 
loss rate, so that in this case detection 
thresholds could be set high enough to cut most of the
electromagnetic background.

Simulations showed that the above
triggers allow the detection of protons and pions with energies
$T_{\mathrm{p}}>$60~MeV and $T_\pi>$50~MeV respectively, 
while electromagnetic background
is suppressed with an efficiency of about 99\%.

The remaining background comes mainly from pair production processes. 
In most of the cases only one of the two electrons enters MIDAS, 
while the other is emitted at a very small angle and then detected 
by the \v{C}erenkov counter (see Figure~\ref{gdh}).
These events were eliminated by requiring an anticoincidence 
between the \v{C}erenkov signal and the MIDAS trigger.
With this configuration the 
background rate
was reduced by about 3 orders of magnitude and kept
comparable to the pion rate. Further suppression was
performed off-line as discussed in Section \ref{ch:pid}.

\subsection{Tracking}\label{ch:tra}

Particle tracking is performed by the V$_1$ and V$_2$ detectors.
A charged particle passing through one of these detectors produces a 
signal from a single
ring and a single sector.

The intersection between the hit ring and the hit sector determines 
the impact point of the particle on the detector. Then two 
detectors allow the complete
reconstruction of the particle trajectory.

The polar and azimuthal angular
resolution depends on the ring pitch and on the sector opening angle
respectively, which were chosen to give 
$\sigma(\vartheta)\approx 1^\circ$ and 
$\sigma(\varphi)\approx 10^\circ$.

\subsection{Particle identification}\label{ch:range}

Most of the protons and pions entering MIDAS have momenta
$p\lesssim$1~GeV/c, so they can be identified by
measuring their \dedx.
In particular we apply the Range Method, previously developed for
DAPHNE \cite{range} and adapted to MIDAS geometry.
A particle entering V$_1$ can cross several layers before stopping or
going through Q$_3$ if its energy is high enough, the signal from
each detector giving the particle \dedx{} along its path.

In the Range Method we use the Bethe-Bloch equation to calculate
the mean energy loss of a given particle type (proton or pion)
in the different MIDAS
layers, as a function of the particle kinetic energy.
Then a Least Squares Fit is used to determine the 
kinetic energy value
which minimize the difference between the measured and calculated 
values of the \dedx{} samples. The analysis of the $\chi^2$ 
distribution of the fit allows the identification
of the particle.

\section{The electronics}

\begin{figure}[hbt]
\begin{center}
\epsfig{file=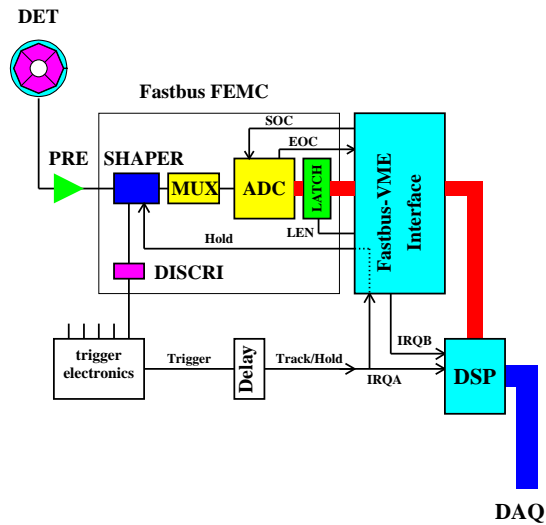, width=8cm}
\end{center}
\caption{Electronic chain}
\label{chain}
\end{figure}

A block diagram of the electronics is shown in Figure~\ref{chain}.
A detailed description can be found in ref.~\cite{midas1}.

The silicon detectors are biased by a CAEN SY527, system remotely
controlled via VME.

The readout electronics is based on Fastbus module FEMC described
in refs.~\cite{delphi1,delphi2}.
These modules receive the
differential signals from the preamplifiers and provide the shaping,
the track \& hold, the multiplexing and the digitization of the 
signals.
There are four FEMC modules used in the electronics set-up,
each module of 80 channels divided into 5 groups (rows) and
each channel is equipped with a hybrid shaper, consisting of a 
(RC)$^2$CR filter.
It is followed by a hold circuit allowing the 80 signals to be 
multiplexed and serially digitized by a single 12-bit ADC.

After the first shaping stage a signal for trigger logic purposes
is output. A double discriminator circuit
allows to set two different threshold levels.

Data from the ADC's are read and decoded into an interface,
which contains control logic and then stored into a buffer memory
under the 
control of a digital signal processor (DSP), Motorola type DSP56001
\cite{agofemc,agodsp,motodsp}.
These data can be transfered to a host computer either via the VME 
bus or via the serial port of the DSP.

The first method, being faster, was used in the experimental
data taking in which a master VMEbus computer handles the readout of the
different sections of the whole apparatus.
The second method was adopted to test MIDAS
independently from the rest of the experimental apparatus.

\section{Tests and performances}

A series of runs at the MAMI tagged-photon beam facility allowed the
 set-up of the entire apparatus.
Photoreactions induced on a Hydrogen target were used
to test the response of the different detector sections.

In particular the response of MIDAS to pions and protons was
studied using the double pion photoproduction reaction
$\gamma \mathrm{p} \rightarrow \mathrm{p} \pi^+ \pi^-$.

A clean sample of events with negligible background
was obtained by a triple coincidence, detecting 2 charged particles
in DAPHNE and one particle in MIDAS. Then two cases are possible:

\begin{enumerate}
  \item the two pions are detected in DAPHNE and the proton in MIDAS
  \item the proton and one pion are detected in DAPHNE and the 
other pion in  MIDAS.
\end{enumerate}

The reaction kinematics is completely defined when the emission 
angles $\vartheta$ and $\varphi$
of both particles detected in DAPHNE are measured, together with 
one of their energies.

In both of the above cases it was possible to calculate precisely
the energy and direction of the third particle by using DAPHNE only,
since this detector has very good angular and momentum resolutions.

\subsection{Angular resolution}

\begin{figure}[hbt]
\begin{center}
\epsfig{file=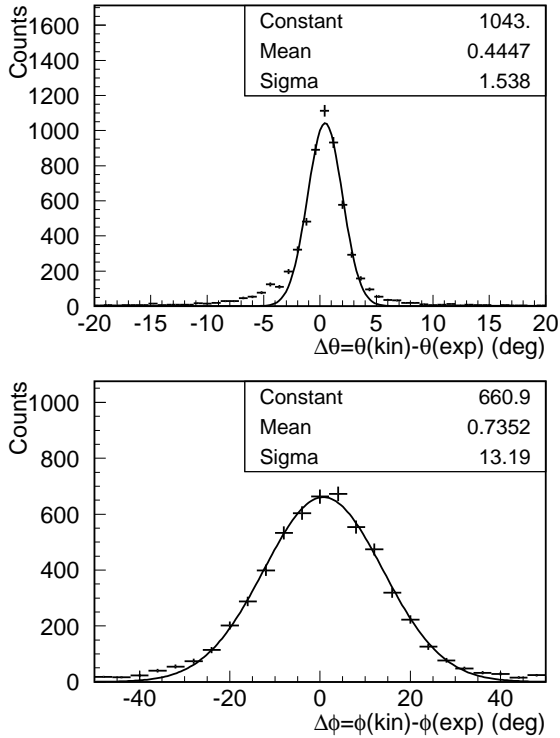, width=11cm}
\end{center}
\caption{Polar (top) and azimuthal (bottom) angular resolutions 
measured using the 
$\gamma \mathrm{p} \rightarrow \mathrm{p} \pi^+ \pi^-$ reaction.}
\label{angres}
\end{figure}

The angular resolution of MIDAS
was obtained by comparing the measured $\vartheta$ and $\varphi$ 
angles of the forward emitted particle with the values 
calculated by kinematics on the basis of the DAPHNE measurements.
The distributions of the difference between these two values
are shown in Figure~\ref{angres}.

The contributions of the uncertainties induced by 
the DAPHNE resolutions
to the distributions widths are $\sigma(\vartheta)\simeq 0.6^\circ$
and $\sigma(\varphi)\simeq 3^\circ$. They can be easily subtracted
to give the overall MIDAS resolution
of $\sigma(\vartheta)\simeq 1.4^\circ$ and
$\sigma(\varphi)\simeq 12^\circ$.

\subsection{Energy calibration and particle identification}
\label{ch:pid}

Most of the selected pions entering MIDAS are relativistic and
release almost the same energy ($E \approx$0.4~MeV) 
in each  silicon detector.
On the contrary  proton energy losses range from
about 0.5~MeV to several MeV, allowing the study of the response
of the detectors at different energies.

The Bethe-Bloch formula was used to calculate the energy released by 
protons, of a given incident energy, in each detector.
The correlation between the pulse height and the calculated energy 
loss determined the detector calibration.

\begin{figure}[htb]
\begin{center}
\epsfig{file=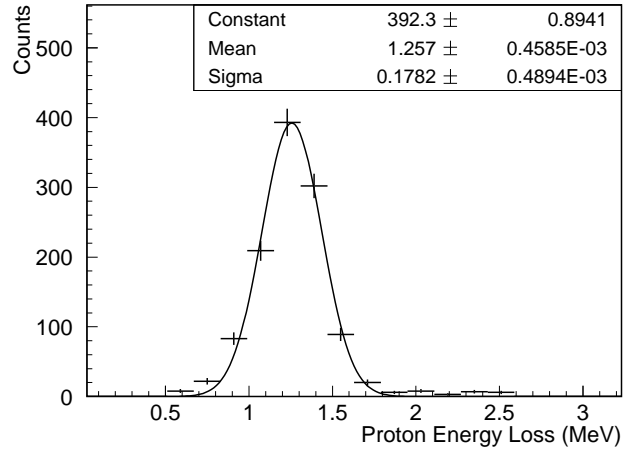, width=9cm}
\end{center}
\caption{Energy loss of $T\simeq$125~MeV protons
measured by Q$_1$ detector. A gaussian fit is superimposed 
on the experimental data.}
\label{prores}
\end{figure}

The response of the Q$_1$ detector
to $T\simeq$125~MeV monoenergetic protons is shown in 
Figure~\ref{prores}.

\begin{figure}[hbt]
\begin{center}
\epsfig{file=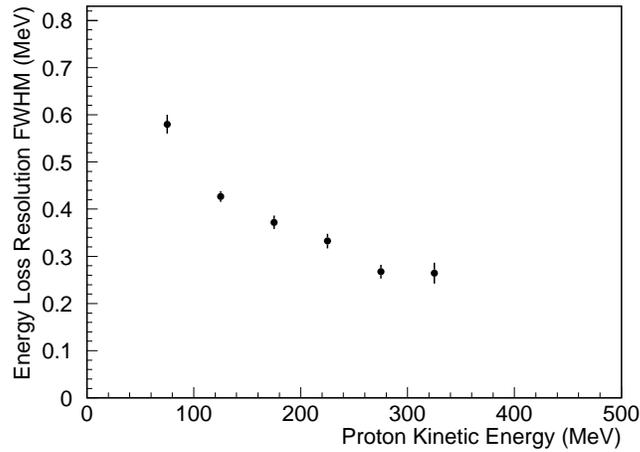, width=9cm}
\end{center}
\caption{The width of the proton energy loss distribution for Q$_1$ 
as a function of the particle energy.}
\label{prores1}
\end{figure}

The mean energy loss is $E\simeq$1.25~MeV, so its
measured distribution width is $\Delta E/E\simeq$34~\% (FWHM).
The energy loss width of the same detector Q$_1$
as a function of the proton energy is shown in 
Figure~\ref{prores1}.

The distribution width is completely dominated by straggling, due
to the small detector thickness. Straggling
is the most serious limitation in
particle identification using \dedx{} techniques.

However MIDAS provides \dedx{} measurements
with enough resolution to discriminate
protons and pions over most of the energy range covered in the
experiment.

\begin{figure}[hbt]
\begin{center}
\epsfig{file=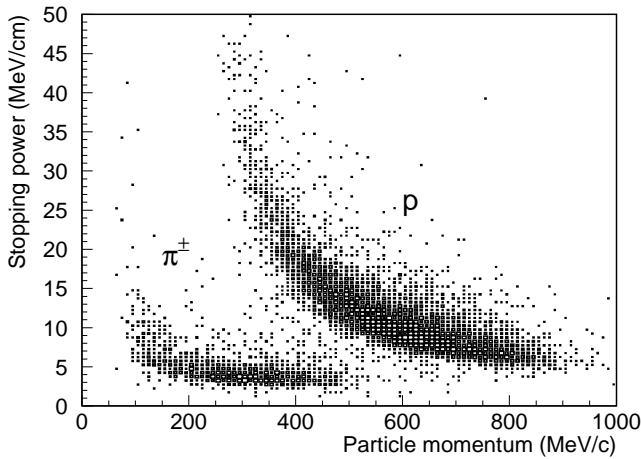, width=9cm}
\end{center}
\caption{Proton and pion stopping power measured in Q$_3$
as a function of the momentum.}
\label{dedx_exp}
\end{figure}

As an example, the stopping power of particles measured
by Q$_3$ is shown in Figure~\ref{dedx_exp} as a function of the 
particle momentum. The pion and proton regions are clearly 
separated and discrimination is possible 
up to a momentum $p\approx 700$ MeV/c.

The Range Method (RM) was used to obtain particle discrimination and
energy reconstruction, as described in Section \ref{ch:range}.
The fit procedure was applied to the selected particles
and a cut on the $\chi^2$ distribution
was established in order to have  good proton identification 
efficiency and minimum contamination of pions.
Pion contamination, estimated by counting pions fulfilling the fit 
condition for protons, is significant only at high energy (3\%
for protons above $p$=500~MeV/c.

\begin{figure}[hbt]
\begin{center}
\epsfig{file=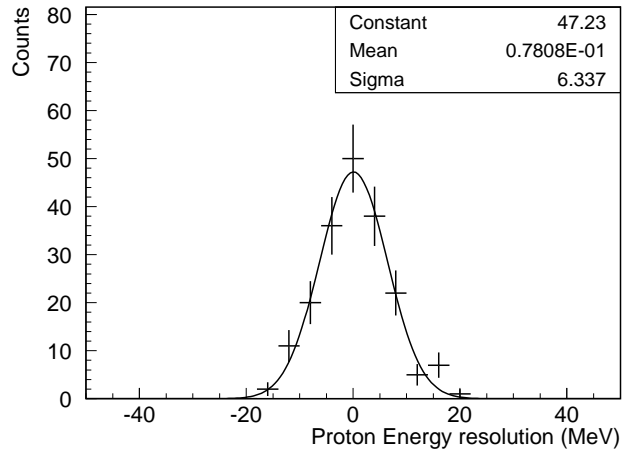, width=9cm}
\end{center}
\caption{Energy resolution of $T\simeq$105 MeV protons}
\label{prorange}
\end{figure}

The proton energy resolution was obtained by comparing the value
calculated by kinematics with that reconstructed
by the RM.
The difference of the two for
protons with kinetic energy around T$\simeq$105 MeV is shown in 
Figure~\ref{prorange}.
The curve is a gaussian fit whose parameters are
reported in the upper box of the plot. The width of the distribution
is $\sigma\simeq$6 MeV which corresponds to a kinetic energy 
resolution $\Delta T/T=$14\% (FWHM).

\begin{figure}[hbt]
\begin{center}
\epsfig{file=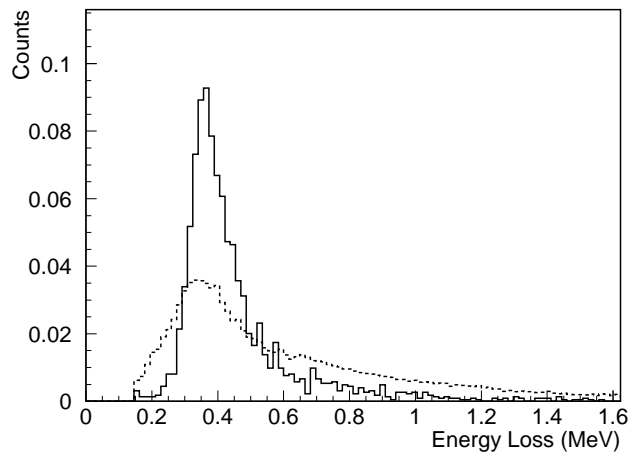, width=9cm}
\end{center}
\caption{Energy loss spectra of pions (continuous line) and
electrons (dashed line) measured in Q$_3$. Counts are normalized to 
give the same total number of events in each spectrum.}
\label{pi-el_ide}
\end{figure}

The pion energy cannot be reconstructed with the RM because
the stopping power of these particles is nearly independent of
energy.
In addition, the separation between pions and the residual 
electromagnetic background, discussed in Section \ref{ch:ht}, cannot 
be fully achieved by \dedx{}. However electrons and positrons, 
producing a shower in the lead absorbers, exhibit an energy loss 
spectrum different from pions, as shown in Figure~\ref{pi-el_ide}.
In this case the electron sample  was obtained by requiring a 
coincidence between the \v{C}erenkov signal and MIDAS, in order to 
detect e$^+$e$^-$ events.

It turns out that we can reject about 40\% 
of electrons while only loosing 10\% 
of pions by selecting particles with  an energy loss 
E$<$0.6 MeV in a single detector. The same condition, applied to all the
detector Q$_1$, Q$_2$ and Q$_3$, produces the suppression of 70\%
of electrons while loosing 20\% of pions.

\section{Conclusions}

The MIDAS device meets the main needs of the GDH experiment for 
which it was build.

The measured angular resolution ($\sigma(\vartheta)\simeq 1.4^\circ$ 
and $\sigma(\varphi)\simeq 12^\circ$) are in reasonable agreement 
with the expected values (see Section \ref{ch:tra})
with discrepancies attributable to multiple scattering.

The Range Method analysis provides a powerful tool to identify 
protons with $<$3\% 
of pion contamination and to measure their energy  with good 
resolution ($\Delta T/T=$14\% 
at T=105 MeV) by the simultaneous use of all the available 
measurements of the particle energy losses.

The e$^\pm$ background was suppressed in the trigger logic by
3 orders of magnitude and further reduced off-line of 70\% 
by applying a cut on the energy loss spectra.
Under these conditions the background contamination to pion count 
rate is still about 50\%. 
This value is acceptable, since the main goal of our experiment is 
the measurement of the hadronic reaction asymmetry and
electromagnetic processes give a net null contribution to it.
On the other hand the background contamination in the small 
angular region of MIDAS does not significantly increase the
statistical error of the measurement.

\begin{ack}
We acknowledge the careful work of the electronic and mechanical
workshops of INFN-sezione di Pavia
 in the realization and assembling of the MIDAS 
set-up. In particular we are indebted with
V. Arena,
G. Bestiani,
E.G. Bonaschi,
D. Calabr\`o, 
A. Freddi, 
G. Iuvino,
P. Ventura,
F. Vercellati.
\end{ack}


\begin{thebibliography}{99}

\bibitem{gdhprop1} J. Ahrens et al., Experimental check of the 
Gerasimov-Drell-Hearn sum rule, {\em MAMI proposal\/} A2/2-93
\bibitem{gdhprop2} J. Ahrens et al., Helicity dependence of single 
pion photoproduction on the proton, {\em MAMI proposal\/} A2/2-95
\bibitem{gdhprop3} J. Ahrens et al., Helicity dependence of double 
pion photoproduction on the proton, {\em MAMI proposal\/} A2/3-95
\bibitem{gdhprop4} J. Ahrens et al., Helicity dependence of single 
and double pion photoproduction and the GDH sum rule on the neutron,
{\em MAMI proposal\/} A2/1-97
\bibitem{daphne} G. Audit et al., Nucl. Instr. and Meth. A301 
(1991), 473
\bibitem{pale} A. Panzeri, {\em Tesi di Laurea\/}, Universit\`a di 
Pavia (1996)
\bibitem{range} A. Braghieri et al., Nucl. Instr. and Meth. 
A343 (1994), 623
\bibitem{midas1} S. Altieri et al, INFN/TC-98/30, 30-10-1998; 
internal report
\bibitem{star} M. Sauer et al., Nucl. Instr. and Meth. A378 (1996), 143
\bibitem{delphi1} G. Barichello et al., Nucl. Instr. and Meth. A254
(1987), 111
\bibitem{delphi2} I. Lippi et al., Nucl. Instr. and Meth. A286 
(1990), 243
\bibitem{agofemc} Acquisizione VME per la scheda di readout FEMC,
INFN sezione di Pavia, Servizio Elettronico; internal report
\bibitem{agodsp} CPU DSP56 and DSPbug 4.1, INFN sezione di Pavia,
Servizio Elettronico; internal report
\bibitem{motodsp} DSP56001 24-bit Digital Signal Processor User's 
Manual, Motorola Inc

\end{thebibliography}
\end{document}